 \documentclass[authoryear,preprint,review,12pt]{elsarticle}

\usepackage{graphics}

\usepackage{graphicx}

\usepackage{epsfig}

\usepackage{amssymb}

\usepackage{amsthm}

\usepackage{lineno}

\journal{New Astronomy}

\begin{document}

\begin{frontmatter}



\title{THE IRREGULAR VARIABLE KU HER: \\
The Light and Period Variations}


\author{Esin Sipahi\corref{cor1}}
\ead{esin.sipahi@mail.ege.edu.tr}
\cortext[cor1]{Corresponding author}

\address{Ege University, Science Faculty, Department of Astronomy and Space Sciences, 35100 Bornova, \.{I}zmir, Turkey}

\begin{abstract}
We present CCD photometric observations of the slow irregular type variable KU\,Her. We have also used observations from the ASAS Database and the AAVSO International Database. The V-band observations span eight years from August of 2002 to September of 2010. The length and density of the data enable us to look for variations on time scales ranging from days to years. The V-band observations of KU\,Her are analysed to derive periodicity. This analysis was an approach to find the fundamental periods of the variability of KU\,Her. Analysis of the KU\,Her light variation indicates a complex combination of different periods. Five main and three harmonic periods are identified. In addition to the period analysis, we present the colour variations of the star.
\end{abstract}


\end{frontmatter}


\section{Introduction}

The long period variables are separated Miras, semiregular and irregular variables in the General Catalogue of Variables Stars \citep{Kho87}. The basic cause of the variability of the light curves of these stars is the pulsation. The Mira variables are the best-defined subgroup, but the semiregular and irregular variables are more numerous. Although the semiregular and irregular variables can reveal important information, as the distinction between these groups is not well defined in the literature. The irregular variables have been almost neglected although their role within the evolution on the AGB and their overall properties are far from being understood. The objects classified as L-type (also called the slow irregular variables) are usually late-type giants (Lb) or supergiant (Lc) for which the variability is associated with pulsation in their atmospheres. There are few works that focused on irregular variables (e.g. \citealt{Ker96}). This situation has changed during the last years for these variables with some published studies (see \citealt {Per09, Per10, Per11}).

According to the General Catalogue of Variable Stars \citep{Kho87}, KU\,Her (IRAS 18235+1210) is a Lb (slow irregular) type variable with light varying from $12^{m}.70$ to $13^{m}.50$. In the SIMBAD database, the magnitude of B is $12^{m}.70$, and its spectral type is given as M3. The light variability of this variable was reported for the first time by \citet{Hof35}. The light curve of KU\,Her was observed by the ASAS project \citep{Poj05} and by AAVSO. According to our knowledge, no period study, nor a spectroscopic and a photometric study of this irregular variable was published so far.

To understand the light variations of irregular variables, the detailed observational material is very much needed. The light changes of these variables typically occur with periods between 30 and 600 days. This makes it very difficult to obtain reliable light curves. In this paper, collecting all available data from various international databases (ASAS, AAVSO) and our observations for KU\,Her, we present the results of period analysis. We briefly discuss the light curve and colour curve variations.

\section{Observations and Light Variations}

KU\,Her was observed at Ege University Observatory in the years 2008 and 2009. The Apogee U47 CCD camera was used for all the observations. All the CCD observations were made with BVR filters, sequentially. We obtained about eight images in each filter for every target star on each observational night. BD +22$^{\circ}$ 4377 and BD +22$^{\circ}$ 4417 were used as the comparison and check stars, respectively. The observational data were transformed into the standard BVR system using SA 109 standard region. We have also used observations from the AAVSO International Database and the ASAS database. The V light and B-V, V-R colour curves of KU\,Her observed in this study are shown in the Figure 1. All the V-band observations (ASAS+AAVSO+this study) are shown in Figure 3 (see Section 3). In these light curves, we used the averaged values for each data set obtained on each observing night. Basic light curve parameters of KU\,Her are given in Table 1.

\begin{figure*}
\hspace{1.6 cm}
\includegraphics[width=23cm]{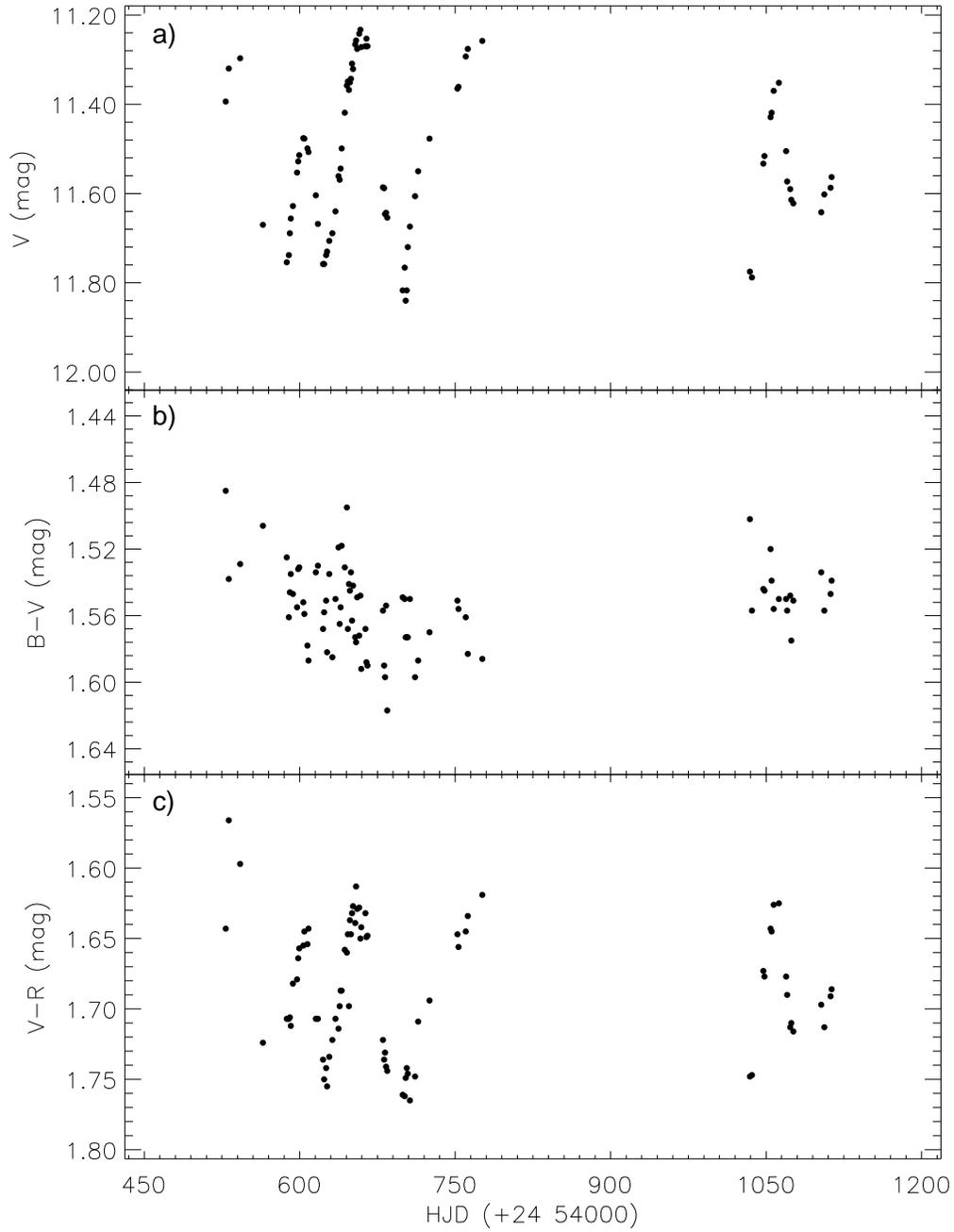}
\vspace{0.2 cm}
\caption{Comparison of V-light curve and B-V, V-R colour variations for KU\,Her.}
\label{Fig.1}
\end{figure*}

The light curve of KU\,Her was very variable in the form, sometimes exhibiting a sharp maximum, sometimes a broad maximum and sometimes an amplitude variation in same observational season. This variable shows a complex LC with a timescale of 60-70 days (see Figure 3). There is also variability on a timescale of thousands of days. In Figure 1 it can be seen that the change in V-R colour resembles the V-light curve. The brightness maxima at HJD 2454658 and HJD 2455062 are accompanied by a maxima in the V-R curve. We present the results as maximum and minimum B-V and V-R values for KU\,Her in Table 2. The V-R colour varies by 1.597 to 1.765 mag for the star. No phase shift was found between the V light and the V-R colour variations. However, there is no variation in the mean B-V colour.

\begin{table*}
\centering
\caption{Basic light curve parameters of KU\,Her.}
\vspace{0.3cm}
\begin{tabular}{ccccccccc}
\hline\hline
Year	&	Time span of	&	\multicolumn{3}{c}{Number of points}	&	\multicolumn{3}{c}{Total Range of the Light Changes}	&	Source\\
	&	light curve	&		&		&		&		&		&		&		\\
\hline																	
	&		&	B	&	V	&	R	&	B	&	V	&	R	&		\\
\hline																	
2002	&	52510-52560	&	-	&	10	&	-	&	-	&	11.4-11.9	&	-	&	1	\\
2003	&	52707-52940	&	-	&	93	&	-	&	-	&	11.0-12.2	&	-	&	1	\\
2004	&	53077-53299	&	-	&	46	&	-	&	-	&	11.1-12.2	&	-	&	1	\\
2005	&	53456-53665	&	-	&	50	&	-	&	-	&	11.0-12.2	&	-	&	1	\\
2006	&	53814-53981	&	-	&	26	&	-	&	-	&	11.0-11.7	&	-	&	1	\\
2007	&	54178-54414	&	-	&	131	&	-	&	-	&	10.8-11.5	&	-	&	1	\\
2008	&	54497-54791	&	65	&	262	&	65	&	12.8-13.4	&	11.0-11.8	&	9.6-10.1	&	1, 2, 3	\\
2009	&	54909-55124	&	17	&	106	&	17	&	12.9-13.3	&	11.2-11.7	&	9.7-10.0	&	2, 3	\\
2010	&	55222-55452	&	-	&	22	&	-	&	-	&	10.9-11.4	&	-	&	2	\\
\hline																	
\end{tabular}																
\begin{list}{}{}															
\item[1]{\small ASAS project}															
\item[2]{\small AAVSO}											
\item[3]{\small This Study}															
\end{list}																	
\end{table*}

\begin{table*}
\centering
\caption{The basic values of KU\,Her light and colour variations.}
\vspace{0.3cm}
\begin{tabular}{cccccc}
\hline\hline
Year	&	$V_{ampl.}$ 	&	$(B-V)_{max}$	&	$(B-V)_{min}$	&	$(V-R)_{max}$	&	$(V-R)_{min}$	\\ 
	&	(mag)	&	(mag)	&	(mag)	&	(mag)	&	(mag)	\\
\hline											
2008	&	11.23-11.84	&	1.51	&	1.60	&	1.60	&	1.77	\\
2009	&	11.35-11.79	&	1.52	&	1.58	&	1.63	&	1.75	\\
\hline											
\end{tabular}
\end{table*}

The infrared magnitudes of the variable are given by \citet{Cut03} as $J=7^{m}.697$, $H=7^{m}.718$ and $K=7^{m}.757$. In Figure 2 we gave two-colour plot from \citet{Leb09}.  KU\,Her is seen in the same region with the irregular variables in this two-colour diagram. The irregular variables are predominantly found in the area of the blue SRVs in agreement with the findings from \citet{Ker94}. 

\begin{figure*}
\hspace{2.2 cm}
\includegraphics[width=15cm]{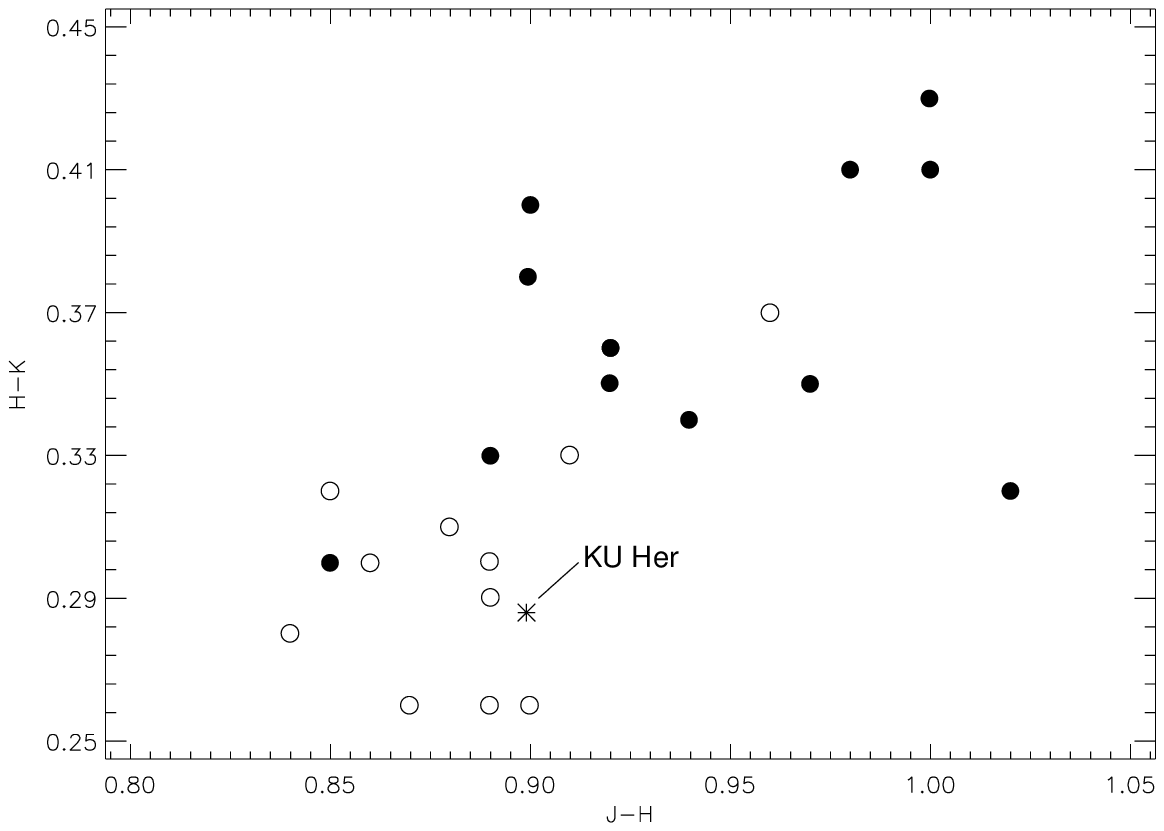}
\vspace{0.2 cm}
\caption{J-H/H-K diagram for the irregular (open circles) and the semiregular (filled circles) variables from \citet{Leb09}. KU\,Her is seen in the same region with the irregular variables.}
\label{Fig.2}
\end{figure*}

\section{The Period Analysis}

Our primary goal in the analysis of the V-light curve of the KU\,Her was to determine the periodic behaviours. This task is complicated by the fact that the cyclic variations of LPV's are not perfectly regular, in fact, they are decidedly variable. Each cycle is likely to have a slightly different period and amplitude, as well as a unique light curve shape. Sometimes the variability amplitude may increase or decrease dramatically, or the period varies significantly from one cycle to the next. Although these factors make the period analysis more difficult, they also give us much information about the nature of these stars.        

The frequency analysis was performed using PERIOD04 \citep{Len05}. The peaks are searched for a wide range from 0 to 0.5 $c d^{-1}$. In the region higher than 0.05, the spectrum always appears in a flat form. The amplitude spectrum of the KU\,Her data is shown in Figure 4. The program gives a signal-to-noise (S/N) value for each detected peak in the amplitude spectrum. The S/N ratios of the resulting frequencies were calculated after prewhitening the corresponding frequency. The applied criterion for the significance of the frequencies was the S/N ratio higher than 4.0 \citep{Bre93, Kus97}. After the period analysis, the eight periods stood out in the V light curve. These peaks are slightly above the noise limit, as four times of the mean noise amplitude \citep{Bre93, Kus97}. 

It is clear from Figure 3 that the behaviour cannot be explained by a single period and is also subject to changes of amplitude and probably the long term variation. In the last few years, the light curve of KU\,Her has shown a fading in the average magnitude and the amplitude. The resulting periods found for KU\,Her are listed in Table 3 together with their S/N ratio. We found that the eight-period curve follows the shape of the V-data. In Figure 3, we show the V-light curve of KU\,Her together with a fit with all periods. This object shows variability with the period of 56.4 days basically, according to the LC. There are also variations in amplitude and mean magnitude on timescale of 459.7 days. We should note that 122.1, 99.6 and 1050.7 days periods are harmonics of each other. 

\begin{figure*}
\hspace{0.7 cm}
\includegraphics[width=12.7 cm]{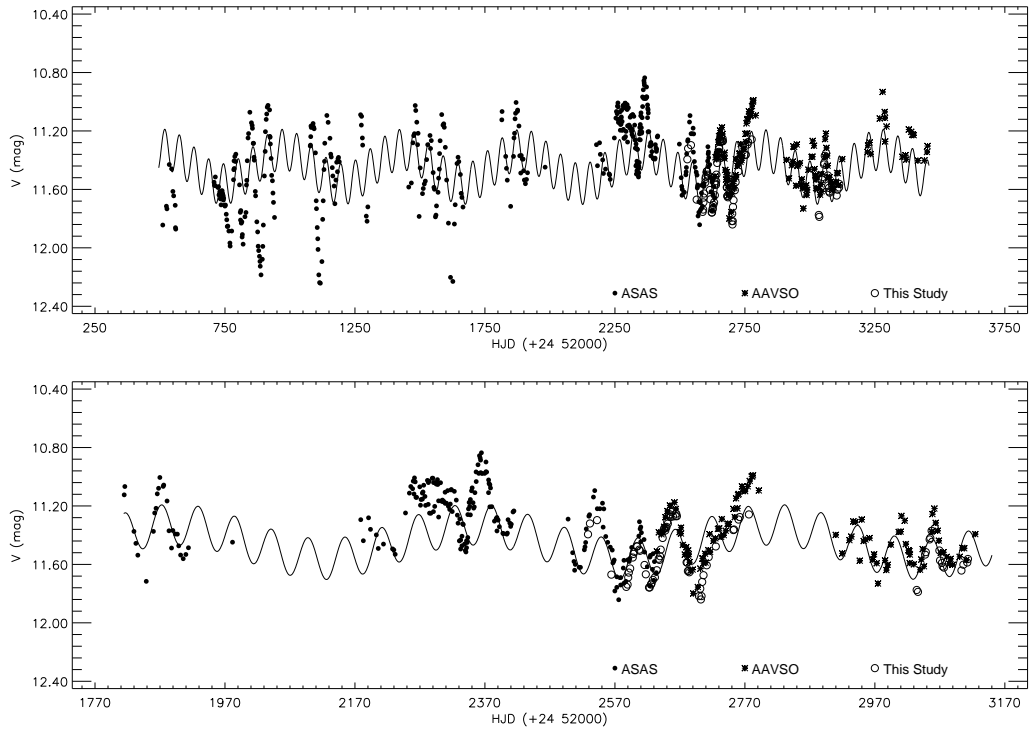}
\vspace{0.2 cm}
\caption{The V-light curve of KU\,Her with the Fourier fit combined with all periods (The light curve obtained between HJD 2453770-2455170 is shown as more detailed in bottom panel for better visibility of light variation).}
\label{Fig.3}
\end{figure*}

\begin{table*}
\centering
\caption{The results of period analysis of KU\,Her.}
\vspace{0.3cm}
\begin{tabular}{cccc}
\hline\hline
	&	Period (day)	&	S/N	\\
\hline							
$P_{1}$	&	459.7	&	7.3	\\
$P_{2}$	&	56.4	&	8.6	\\
$P_{3}$	&	3677.5	&	6.3	\\
$P_{4}$	&	60.4	&	7.1	\\
$P_{5}$	&	122.1	&	6.2	\\
$P_{6}$	&	99.6	&	4.6	\\
$P_{7}$	&	1050.7	&	5.1	\\
$P_{8}$	&	63.5	&	4.5	\\
\hline							
\end{tabular}
\end{table*}

\begin{figure*}
\hspace{2 cm}
\includegraphics[width=13.2 cm]{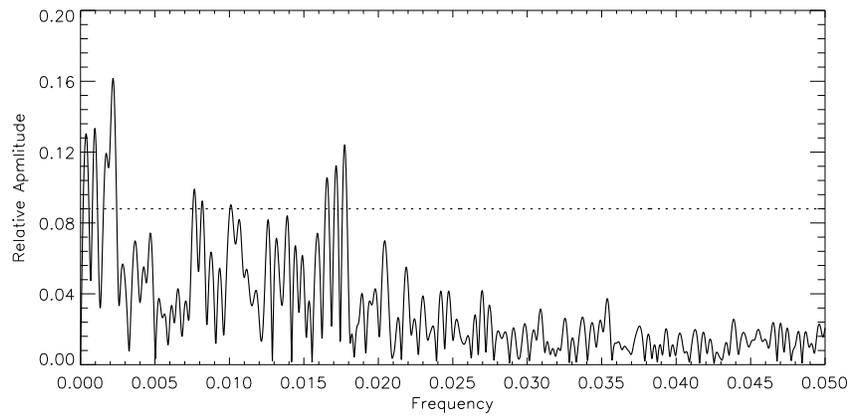}
\vspace{0.2 cm}
\caption{Normalized spectrum of KU\,Her's V-light curve.}
\label{Fig.4}
\end{figure*}

\section{The Results}

The photometric results of the irregular variable KU\,Her are summarized as following:

- KU\,Her shows both the long-term and the short-term brightness and colour variations. The most interesting thing is the gradual fading of KU\,Her's mean magnitude in the last years. This is easily seen in Figure 3. This may be an indicator of the long term evolutionary changes. The V-R colour resembles the V light curve. The brightness maxima in V is accompanied by a maxima in the V-R curve. The same case is also seen in V-I colour for the other stars \citep{Per01}. \citet{Leb09} found a well expressed relation between V and V-I. 

- The complete set of data suggests that the V-light curve of KU\,Her can be qualitatively represented with five periods and three harmonics. However, the problem of the period analysis of such star is to attribute a physical meaning to such a large number of periods. While the longest periods found in the analysis are very unlikely due to pulsation (e.g. \citealt{Woo00}), the shorter periods possibly represent different pulsation modes. In this analysis, some periods are close to each other. These periods could be due to a visual period and a second period, which lags behind the previous, slowing by the stellar envelope. The energy pulse comes up from the core of the star to the surface. In this case, a residual pulse attenuated by the matter, in which the pulse passes through. Therefore, this pulse causes a second period, slightly longer one. This is similar and may be related to what is known as the "echo phenomenon" or "overtones" (see \citealt{Fam92}). The long-term variations of the red supergiants were due to "the convective turnover time of giant convection cells in the stellar envelope" was described by \citet{Stl71}. This process could cause the harmonics. If the matter inside the convection cell is not completely cycled in one period, but the matter clumps and is moved through in three or four periods. 

- The observations of KU\,Her should be continued. This is supported by the fact that for a number of stars originally classified as the irregular variables, a periodicity could later be found better light curve data or an improved data reduction (e.g. \citealt{Per96}). The reason for the irregularity in the light change observed in Lb variables has not been identified yet. Such studies require much larger photometric time series of these stars as it is presented by this paper.

\section*{Acknowledgments} I thank to Ege University Observatory for providing observing time and also thank the AAVSO observers and headquarters staff. I wish to thank the referee for useful comments that have contributed to the improvement of the paper.


\begin{thebibliography}{20}
\bibitem[Breger et al.(1993)]{Bre93} Breger, M. et al., 1993, A\&A, 271, 482 
\bibitem[Cutri et al.(2003)]{Cut03} Cutri R. M., et al., 2003, The IRSA 2MASS All-Sky Point Source Catalog, NASA/IPAC Infrared Science Archive (http://irsa.ipac.caltech.edu/applications/Gator/)
\bibitem[Fadeyev \& Muthsam(1992)]{Fam92} Fadeyev, Y.A., Muthsam, H., 1992, A\&A, 260, 195
\bibitem[Foster(1995)]{Fos95} Foster, G., 1995, AJ, 109, 1889
\bibitem[Hoffmeister(1935)]{Hof35} Hoffmeister, C., 1935, AN, 255, 401
\bibitem[Kerchbaum \& Hron(1994)]{Ker94} Kerchbaum, F., Hron, J., 1994, A\&A Suppl. Ser., 106, 397
\bibitem[Kerchbaum et al.(1996)]{Ker96} Kerschbaum, F., Lazaro, C., Habison, P., 1996, A\&A Suppl. Ser., 118, 397
\bibitem[Kholopov et al.(1987)]{Kho87} Kholopov, P. N., et al. 1987, General Catalogue of Variable Stars Vol. III (4th ed.; Moscow: Nauka)
\bibitem[Kusching et al.(1997)]{Kus97} Kusching, R., Weiss, W.W., Gruber, R., Bely, P., Jenkner, H., 1997, A\&A, 328, 544
\bibitem[Lenz and Breger(2005)]{Len05} Lenz, P., Breger, M., 2005. Commun. Asteroseismol. 146, 53
\bibitem[Lebzelter \& Obbrugger(2009)]{Leb09} Lebzelter, T., Obbrugger, M., 2009, AN, 330, 390
\bibitem[Percy et al.(1996)]{Per96} Percy, J.R., Desjardins, A., Yu, L., Landis, H., 1996, PASP, 108, 139
\bibitem[Percy et al.(2001)]{Per01} Percy, J.R., Wilson, J.B., Henry, G.W., 2001, PASP, 113, 983
\bibitem[Percy et al.(2009)]{Per09} Percy, J.R., Esteves, S., Lin, A., Menezes, C., Wu, S., 2009, JAAVSO, Vol.37, 71
\bibitem[Percy \& Long(2010)]{Per10} Percy, J.R., Long, J., 2010,  JAAVSO, Vol.38, 161
\bibitem[Percy \& Terziev(2011)]{Per11} Percy, J.R., Terziev, E., 2011, JAAVSO, Vol.39, 1
\bibitem[Pojmanski et al.(2005)]{Poj05} Pojmanski, G., Pilecki, B., Szczygiel, D., 2005. AcA 55, 275
\bibitem[Stothers \& Leung(1971)]{Stl71} Stothers, R., Leung, K., 1971, A\&A, 10, 290
\bibitem[Wood(2000)]{Woo00} Wood, P.R., 2000, PASA, Vol.17, 18
\end{thebibliography}
\end{document}